\documentclass[12pt]{article}
\usepackage{epstopdf}
\usepackage[english]{babel}

\usepackage{graphicx}
\usepackage{dcolumn}
\usepackage{bm}
\usepackage[OT1]{fontenc}
\usepackage{amsmath}
\usepackage{amsfonts}
\usepackage{url}

\begin{document}
\begin{center}
\large{
\textbf{The n-body problem - an alternative scheme for determining solutions for planetary systems}
}  
\end{center}

\vspace{1cm}
\normalsize
\begin{center}
    Pawe\l{} Wojda\\
\end{center}
\small
Email: pawwojda@pg.edu.pl\\
Gdansk University of Technology, \\
Faculty of Applied Physics and Mathematics,\\
Institute of Applied Mathematics, 80-233 Gdansk, Poland
\\
\\
\author{
\name{Pawe\l{} Wojda
\thanks{CONTACT P. Wojda. Email: pawwojda@pg.edu.pl}}
{Gdansk University of Technology, 
Institute of Applied Mathematics, 80-233 Gdansk, Poland}
}
\normalsize
\begin{abstract}
This study presents a general alternative scheme of the procedure and necessary conditions for solving the n-body problem. The presented solution is not a solution of the classical problem, where the initial conditions of positions and initial velocities/momenta of the bodies are known. 
Starting from the standard initial condition the procedure treats contributions to momentum from particular pair-wise interactions as independent variables from which the total momentum and velocity of a given mass is then reproduced. Initial values of those contributions are arbitrary as long as the resulting velocities match the initial condition. 
The obtained solutions take into account the gravitational interactions between each pair of bodies, as a result of which they are characterized by higher stability than the solutions of the classical problem.
 
The presented procedure was used to calculate the positions and mutual velocities of three bodies - the Sun, the Earth and the Moon. It has also been tested for our entire planetary system (8 planets) with the Sun and the Moon. For these types of systems, the method allows obtaining solutions that are stable. The procedure was also tested on the example of the Pythagorean system of bodies.

\end{abstract}

{keywords:}\\
\textit{n body problem; Newton's law of universal gravitation}

\section{\label{introduction} Introduction}
According to the law of universal gravitation, any two bodies attract each other with a force that is directly proportional to the masses of the bodies and inversely proportional to the square of the distance between them. For any two bodies, we can therefore write a differential equation describing the change in momentum:
$$\frac{d p_1}{dt}=\frac{G m_1 m_2}{\|r_1-r_2\|^3}(r_2-r_1),$$
$$\frac{d p_2}{dt}=\frac{G m_1 m_2}{\|r_1-r_2\|^3}(r_1-r_2),$$
where $p_i=(p_{xi},p_{yi},p_{zi})$ denotes the momentum of body $i$, $r_i=(x_i,y_i,z_i)$ its position, $m_i$ - mass, and $G$ is the gravitational constant. The position and momentum of a body depend on time.
Given initial conditions for positions and velocities, Newton derived analytical solutions to this problem. He also established in \textit{Philosophi{\ae} Naturalis Principia Mathematica} \cite{Newton} that to solve a problem in which three bodies interact, one must solve a system of differential equations. From the mathematical point of view, the problem of three bodies interacting gravitationally comes down to solving a system of equations for positions and momenta:
$$\frac{d p_1}{dt}=\frac{G m_1 m_2}{\|r_1-r_2\|^3}(r_2-r_1)+\frac{G m_1 m_3}{\|r_1-r_3\|^3}(r_3-r_1),$$
$$\frac{d p_2}{dt}=\frac{G m_2 m_1}{\|r_1-r_2\|^3}(r_1-r_2)+\frac{G m_2 m_3}{\|r_2-r_3\|^3}(r_3-r_2),$$
$$\frac{d p_3}{dt}=\frac{G m_3 m_1}{\|r_1-r_3\|^3}(r_1-r_3)+\frac{G m_3 m_2}{\|r_2-r_3\|^3}(r_2-r_3),$$
$$ \frac{d r_1}{dt}=v_1,$$
$$ \frac{d r_2}{dt}=v_2,$$
$$ \frac{d r_3}{dt}=v_3.$$
Since then, many physicists and mathematicians have tried to solve the three-body problem, i.e. to determine solutions to 18 first-order differential equations with initial conditions (for positions and momenta) or, respectively, 9 second-order differential equations, with some success. Euler, Lagrange, and Poincar\'{e}, among others, have been interested in solving this problem \cite{Euler, Lagrange, Poincare1, Poincare2}. Poincar\'e showed that the three-body problem is not integrable. In other words, the general solution to the three-body problem cannot be expressed in terms of algebraic and transcendental functions through unique coordinates and velocities of the bodies \cite{Poincare1, Poincare2}. Despite the indicated work of Poincar\'{e}, the interest in finding solutions to this problem is still relevant. 
This is evidenced by articles published in recent years on chaotic, periodic or mathematically reliable (within a specified time interval) numerical solutions of three-body problem \cite{Boekholt, Liao1, Liao2, Breen, Suvakov, Li1, Li2, Li3, Hristov}.  In order to find solutions to this system, certain rules are imposed, such as calculations with respect to the center of mass. The influence of angular momentum on the solutions is also considered, attempts are made to reduce the number of equations, and an island of ,,stability'' is sought within which the system will be characterized by periodicity.

In this paper we will present a slightly different approach to obtaining a solution to the three-body problem and more generally the n-body problem. From the principle of relativity we know that the velocities of objects depend on the reference frame, so providing one reference frame, for example the center of mass, significantly narrows the applicability of solutions. 
Since we are interested in the mutual motion of objects, we can assume that we will be interested in the mutual interactions of each body with the other bodies. We do not have to be attached to any frame of reference.
To connect each body with the other bodies, in the presented approach, for each body we should know its total momentum (or velocity) resulting from its mutual gravitational interaction with the other bodies, which means that for a system of $n$ bodies, apart from the initial positions, we should provide $3(n^2 -n)$ initial values of velocities or partial momenta. The initial values of the contributions of partial momenta/velocities can be arbitrary, as long as the sum of these momenta/velocities corresponds to the initial conditions, i.e. the total momenta for each body.  
In total, this gives us $3n^2$ initial conditions. However, the question remains what equations will be responsible for the evolution of these initial conditions in time. The answer to this question has already been included in this paper. The idea that any two bodies interact was already expressed by Newton \cite{Newton}. We will get additional $3 (n^2 - n)$ equations by adding to the equations of $n$ bodies the interactions between each pair of bodies:
$$\frac{d p_{ij}}{dt}=\frac{G m_i m_j}{\|r_i-r_j\|^3}(r_j-r_i)=F_{ij},$$
where $p_{ij}$ is the momentum that a body of mass $m_i$ generates under the influence of the gravitational interaction of mass $m_j$.

In this way we obtain a system of $3 (n^2 +n)$, where $3n$ equations are equations for positions and $3n$ equations are commonly used equations for the change of momentum. Of these, for the three bodies, 9 equations for momentum and 9 equations for position are equivalent to the equations given by Newton for the 3-body problem. In general, for $n$ bodies these equations will take the form:
$$\frac{d p_{ii}}{dt}= \sum_{j\neq i} \frac{G m_i m_j}{\|r_i-r_j\|^3}(r_j-r_i)=F_{ii},$$
$$ \frac{d r_i}{dt}=v_i.$$

Given $3(n^2 +n)$ equations with $3n^2$ initial conditions, it may seem that the system has $3n$ more differential equations than initial conditions, however it can be seen that the total momentum of body $i$ should equal its sum of partial momenta $p_{ii}=\sum_{j\neq i} p_{ij}$. Thanks to this, we have the same number of equations as initial conditions.
The problem of momentum change can be written with a single matrix equation:
$$\frac{d \mathbf{P}}{dt} = \mathbf{F},$$
where $\mathbf{P}$ is the momentum matrix with elements $P_{ij} = p_{ij}$ and $\mathbf{F}$ is the gravitational interaction force matrix with elements $F_{ij}$.

It remains therefore to combine these equations into one coherent whole. We will search for solutions to the system using numerical integration. The system of differential equations preserves energy:
$$E= \sum_{i} \left(\frac{P_{ii}^2}{2 m_i} - \sum_{j>i} \frac{G m_i m_j}{\|r_i-r_j\|}\right),$$
so also the solution obtained by the numerical scheme must satisfy this equation.

From the energy equation we can derive the Hamilton equations:
$$\frac{d E}{d P_{ii}}  = v_i = \frac{d r_i}{d t},$$
$$\frac{d E}{d r_i}  = - F_{ii} = - \frac{d P_{ii}}{d t}.$$
Moreover, from the energy equation we can derive additional relations, which describe gravitational interaction between each pair of bodies (for $j$ different from $i$):
$$\frac{d E}{d (r_i-r_j)}=-F_{ij}=-\frac{d P_{ij}}{dt}.$$
The numerical scheme must also take into account that for those additional equations  the following equality holds:
$$\sum_{j\neq i} \frac{P_{ij}}{m_i}=v_i=\frac{d r_i}{dt}. \;\;\;(*)$$

\section{Numerical scheme}
To solve the Cauchy problem:
$$\frac{d x(t)}{dt} = f(t, x(t)), \qquad x(t_0)=x_0$$
we will use the Runge-Kutta method of the second order:
$$x(t_{k+1/2})= x(t_k)+\frac{\tau}{2} f(t_k, x(t_k)),$$
$$x(t_{k+1})= x(t_k)+\tau f(t_{k+1/2}, x(t_{k+1/2})).$$
The Runge-Kutta method for the half-step will be applied to the elements that do not lie on the diagonal of the matrix $d \mathbf{P}/{dt}$:
$$P_{ij}^{k+1/2}= P_{ij}^k + \frac{\tau}{2} F_{ij}^k,$$
then the solution for the momentum $P_{ii}^{k+1/2}$ will not be approximated by numerical methods, but instead the relation will be used $P_{ii}^{k+1/2} = \sum_{j\neq i} P_{ij}^{k+1/2}$. This equality can tell us what momentum the body 
$i$ has with respect to the other bodies. In this way, the solution for $P_{ii}^{k+1/2}$ will be coupled to the solutions for partial momenta with the same first index $i$.
Based on the values of momentum, we can determine the velocities of objects. In the case of Newtonian mechanics, the velocity will be:
$$v_i = \frac{P_{ii}}{m_i}.$$
Having the velocity, we determine the change in the position of individual bodies:
$$r_i^{k+1/2} = r_i^k + \frac{\tau}{2} v_i^k.$$
Thanks to the last formula, we can determine the forces $F_{ij}$ acting on the body in the half-step $k+1/2$. Then we determine the change of all momenta of the matrix $\mathbf{P}$:
$$P_{ij}^{k+1} = P_{ij}^k + \tau F_{ij}^{k+1/2},$$
and then we determine the velocities and changes in the positions of the bodies:
$$r_{i}^{k+1} = r_{i}^{k} + \tau v_{i}^{k+1/2}.$$

In this way, we are able to unambiguously go from the positions and momenta at the kth time moment to the positions and momenta at the $k+1$st time moment. The presented scheme allows us to obtain a solution that satisfies the principle of conservation of energy. Moreover, this scheme allows us to represent for each body the relationship given by equations (*), which allows the additional equations for each pair of bodies to satisfy the standard equation of conservation of energy.  
In order for the given solution to be numerically stable, a time step should be selected that will regulate the magnitude of changes in momenta and positions with respect to the initial values:
$\tau\ll m_i \|r_{ij}\|^2/(F_{ij}\cdot r_{ij})$, for each $j$ different from $i$, where $r_{ij}= r_j - r_i$.

The presented scheme allows for obtaining solutions of equations describing the motion of bodies in a simple way, but it does not fully demonstrate the advantages of including additional equations. Similar solutions can be obtained by solving standard Newton equations describing motion caused by the gravitational interaction of bodies with appropriately changed initial conditions, in which the initial velocity is replaced by the velocity resulting from the total gravitational interaction of bodies with all other bodies. However, the properties of the solutions of the presented equations can be significantly improved by using leap-frog type methods:
$$x(t_{k+1/2})= x(t_{k-1/2})+\tau f_1(t_k, y(t_k)),$$
$$y(t_{k+1})= y(t_k)+\tau f_2(t_{k+1/2}, x(t_{k+1/2})).$$
By determining solutions of additional equations for momentum at half-steps one can significantly improve the stability of the obtained solutions.
The numerical solution for the half-step moment can be obtained using the following scheme:
$$P_{ij}^{k+1/2}= P_{ij}^{k-1/2} + \tau F_{ij}^k,$$
$$P_{ii}^{k+1/2} = \sum_{j\neq i} P_{ij}^{k+1/2},$$
$$v_i^{k+1/2} = \frac{P_{ii}^{k+1/2}}{m_i}.$$
And for the total-step moment:
$$r_{i}^{k+1} = r_i^k + \tau v_i^{k+1/2}.$$

Despite the necessity of determining solutions of $3(n^2+n)$ equations, the computational complexity, due to the necessity of determining forces on the diagonals of the matrix $\mathbf{F}$, is proportional to the number of operations performed when solving the classical problem.

\section{Results}
\subsection{Sun - Earth - Moon}
The procedure was tested to determine the positions and mutual velocities resulting from gravitational interaction of the Sun-Earth-Moon system. This problem is interesting because we observe it every day, but in most cases we divide it into two problems - the orbit of the Earth around the Sun and the orbit of the Moon around the Earth, i.e. we solve two separate problems. Each of them behaves stably, but when they are combined into one whole, the Moon most often escapes the Earth's influence, and with slightly ,,better'' selected initial conditions, the Moon collides with the Earth. Reducing the time step does not allow to fix this problem. Finding a solution to this problem is mundane, and at the same time it stimulates the imagination. 
Both procedures presented in the article were used to determine the solutions of the problem under different initial conditions. The solutions obtained using the first numerical scheme allowed obtaining solutions over a period of tens of thousands of years. Ultimately, however, the solutions lost their accuracy leading to a loss of stability. Using the leap-frog scheme, the problem of the gravitational interaction of the Sun - Earth - Moon was tested for a time period of up to 5 million years, with the solutions remaining stable for the entire period studied.

Example graphs of solutions obtained with the following initial conditions for:
\begin{itemize}
    \item Earth:\\
$m_E = 5.97219\cdot 10^{24}$ kg,\\
 $r_1 = (1.52098\cdot 10^{11}, 0 , 0)$m,\\
 $v_{1S}=(0, 29290, 0)$m/s,
$v_{1M}=(0, 0, 0)$m/s, 
\item Moon:\\
$m_M = 7.347673\cdot 10^{22}$ kg,\\
 $r_2 =  r_1 - (\cos(5 \pi/180)\cdot 4.05696\cdot 10^8, 0, - \sin(5 \pi/180) \cdot 4.05696\cdot 10^8)$m,\\
$v_{2E} = (0, -970, 0)$m/s,
$v_{2S} = (0, 29288, 0)$m/s, 
\item Sun:\\
$m_S = 1.98855\cdot 10^{30}$ kg,\\
$r_3 = (0, 0, 0)$ m,\\
$v_{3E} = (0, 0, 0)$m/s,
$v_{3M} = (0, 0, 0)$m/s, 
\end{itemize}
obtained with a time step of $1/10000$ of an Earth year (about 52.6 minutes), where a year is assumed to be 365.256363 days, are listed below. Although the procedure requires more initial conditions, as can be seen, they are in line with intuition and can be found in the publicly available literature. When comparing the positions of the Sun-Earth-Moon system obtained for a constant ratio of two time steps, the difference in solutions becomes smaller with decreasing time step - this proves the stability of the method.

Figure 1 shows the dependence of the change in the distance from the Moon to the Earth over the course of one year. The obtained results are qualitatively consistent with the publicly available results \cite{Meeus}. The graph shows the repeatability of the solution for the distance between the Earth and the Moon.

Figure 2 shows the positions of the Moon relative to the Earth over a period of 10 years. The Moon's orbit relative to the Earth takes place in irregular ellipses. The inclination of the Moon's rotation axis to the ecliptic plane, which is the assumed ,,plane'' of the Earth's movement relative to the Sun, causes the Moon not to move within one plane but is spatial. The rotation of the Moon's orbit can be observed on the graph.

The given formulation of the problem allows us to learn about the momentum that each body has in relation to every other. On this basis, it is possible to know/reconstruct the velocities of bodies resulting from their gravitational interaction.  
Figure 3 shows the change in the speed of the Moon's rotation relative to the Earth. One can notice the periodicity of the changes in speed over time. Based on the results of the calculations of the case discussed in the article, there is also a change in the speed of the Earth relative to the Moon. This speed ranges from 0 to 25 m/s.

Figure 4 shows the annual changes in the position of the Earth and the Moon relative to the Sun. From the graph, it can be seen that the proper motion of the Earth and the Moon is not an ellipse but a spiral. The Earth's motion relative to the Sun is counterclockwise. The Moon orbits the Sun along with the Earth. The Moon's motion relative to the Earth is also counterclockwise.

\begin{figure*}[htb]
{
       \includegraphics[width=0.98\textwidth]{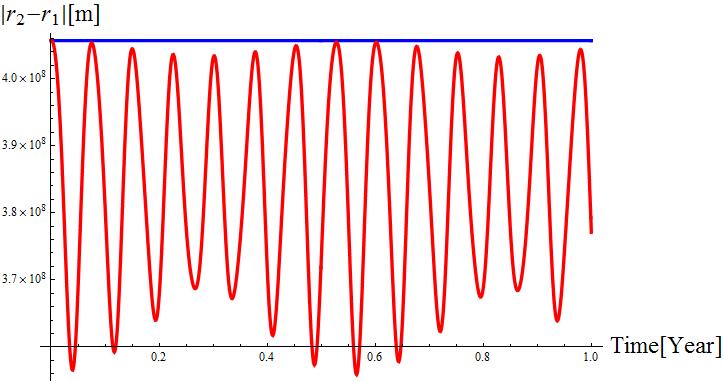}
          
          }            
    \caption{Change in the distance of the Moon from Earth (red line) over one year and the initial distance of the Moon from Earth (blue line).}
    \label{fig1}
\end{figure*}

\begin{figure*}[hptb]
{
       \includegraphics[width=0.98\textwidth]{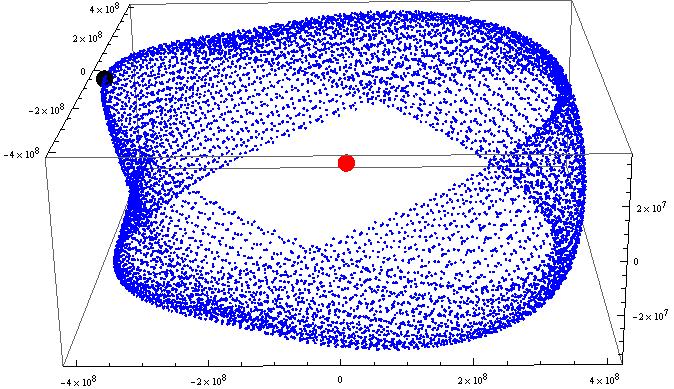}
          
          }            
    \caption{Variation of the Moon's position relative to the Earth (marked with a red dot) obtained over ten years in relation to the Moon's initial position (marked with a black dot). The Moon orbits the Earth counterclockwise.}
    \label{fig2}
\end{figure*}

\begin{figure*}[hptb]
{
       \includegraphics[width=0.98\textwidth]{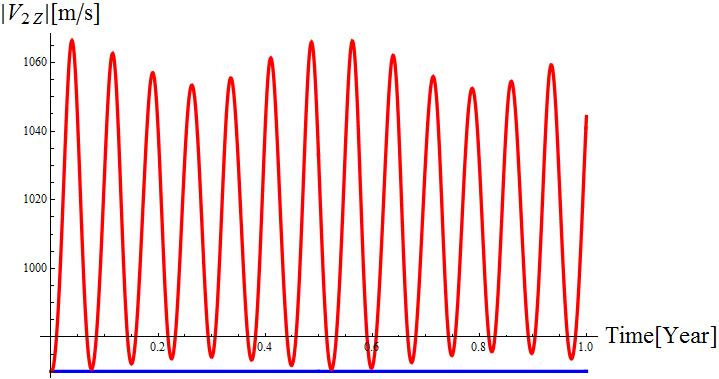}
          
          }            
    \caption{Change in the Moon's velocity relative to Earth (red line) over one year and the Moon's initial velocity relative to Earth (blue line).}
    \label{fig3}
\end{figure*}

\begin{figure*}[hptb]
{
       \includegraphics[width=0.98\textwidth]{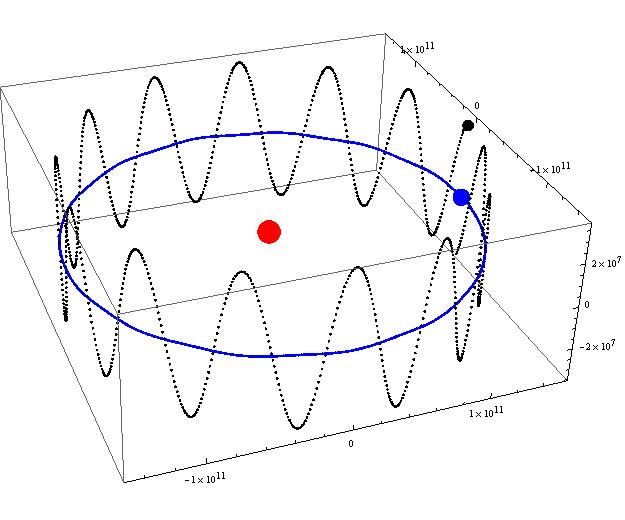}
          
          }            
    \caption{Variation of the position of the Earth (marked by small blue dots) and the Moon (marked by small black dots) relative to the Sun (marked by a large red dot) obtained over the course of one year relative to the initial position of the Earth (marked by a large blue dot) and the Moon (marked by a medium-sized black dot). From this perspective, the Earth orbits the Sun counterclockwise. The axes shown on the graph have a scale expressed in meters. 
}
    \label{fig4}
\end{figure*}

The only points distinguished for obtaining solutions were the positions of the bodies. No assumptions were made regarding the considerations of the center of mass or the conservation of angular momentum. Such assumptions would limit the variety of solutions obtained, since as can be seen, the motion does not take place in one specific plane.
\subsection{Planetary system with Sun and Moon}
The above results show qualitative agreement with observed data. To check the accuracy of the procedure, it was also tested for the entire planetary system, that is, all eight planets along with the Sun and the Moon. Data on positions and velocities were obtained from the Horizons system \cite{Horizons}.  
The model uses the positions of the bodies and the velocities of the bodies relative to the common center of mass on March 27, 2025. Additionally, the velocities of the planets relative to the Sun and the velocities of the Moon relative to the Earth were also used. 
The velocity of the Moon relative to the Sun was assumed to be equal to the velocity of the Earth relative to the Sun. 
The remaining velocities resulting from the gravitational interaction between the bodies were assumed as zero. The leap-frog method was used for the test. Analysis of the solutions was carried out for different time periods. The maximum period was half a million years. During the tests, the Earth and the Moon remained in a common gravitational interaction, which indicates the numerical stability of the solutions.
The accuracy of the solutions was checked based on the analysis of the Earth's positions relative to the Sun after 10 and 100 years obtained using the presented procedure and the results from the Horizons system. The relative error of the results after 10 years was 1.8 percent, and after 100 years - 11 percent. The accuracy of the Moon's positions relative to the Earth was estimated based on a comparison of the number of minimum and maximum distances (perigee and apogee number) of the Moon from the Earth in a period of 100 years and 200 years. In both cases the apogee number was exact (assuming that the anomalistic month (from perigee to perigee) is 27.55455 days). Over the 200-year period, the perigee number was 1 less than the correct number and for 100-year period - it was exact, which means that the relative error in the perigee number was less then 0.04 percent. 
Considering that the assumed velocity of the Moon resulting from its gravitational interaction with the Sun was approximated by the velocity of the Earth relative to the Sun, and with the available data it is difficult to estimate this velocity precisely, 
the accuracy of the obtained results can be assessed as good. The accuracy of the results can be improved by improving the accuracy of the velocity of the Moon, which results from its gravitational interaction with the Sun.

\subsection{Pythagorean body problem}

In the case where the mass ratio of the interacting bodies is comparable, as for example in the Pythagorean case $m_1=5$, $m_2=4$, $m_3=3$, $\|r_1-r_2\|=3$, $\|r_1-r_3\|=4$, $\|r_2-r_3\|=5$ \cite{ Burrau}, the procedure may require a very small time step to maintain stability. In such a case it would be advisable to replace it with a more efficient leap-frog type of method.

Below are the considerations for the case of the Pythagorean system with constant $G=1$:
\begin{itemize}
    \item $m_1 = 5$, $r_1 = (-1,-1,0)$, $v_1 = (0,0,0)$,
\item	$m_2 = 4$, $r_2 = (2,-1,0)$, $v_2 = (0,0,0)$,
\item	$m_3 = 3$, $r_3 = (-1,3,0)$, $v_3 = (0,0,0)$.
\end{itemize}
In order not to modify the original problem, all relative velocities resulting from the gravitational interaction of each pair of bodies were assumed to be zero. The calculations were performed for different time steps. In the performed calculations it was noticed that determining a stable trajectory of bodies for times smaller than 10 time units does not require high computational costs. However, with subsequent interactions of bodies, their mutual approach causes changes in quantities in the time moments following the approach, which have an impact on the stable determination of the trajectory. The approach, which occurs around the 16th time unit, is particularly sensitive to changes in time steps. After this approach of the bodies, one can observe chaotic behavior of the system, which changes depending on the adopted time step. Determination of a trajectory after the 60th time unit is possible using the presented algorithm, but requires time steps of the order of $10^{-8}$ or smaller. This is an advantage that other currently used methods do not allow \cite {Boekholt}. Figure 5 shows the energy values as a function of time. Figure 6 show the trajectories of individual bodies from the Pythagorean system.

All calculations were performed on double-precision floating-point numbers, so the accuracy can be improved, among others, by using the methods proposed in \cite{Burrau, Liao1, Liao2}.
\begin{figure*}[hptb]
{
       \includegraphics[width=0.98\textwidth]{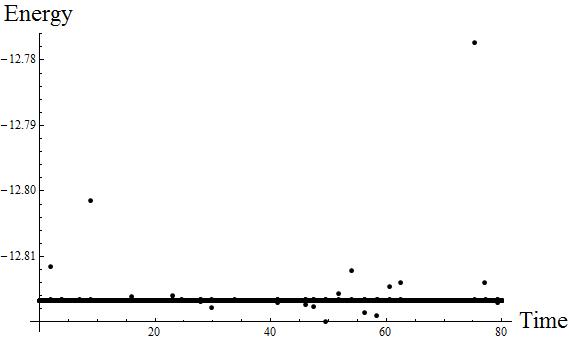}
          
          }            
    \caption{The graph shows the energy for the Pythagorean system of bodies at a time step of $10^{-8}$ (sampled with a step of 0.008).
}
    \label{fig5}
\end{figure*}

\begin{figure*}[hptb]
{
       {\includegraphics[width=0.45\textwidth]{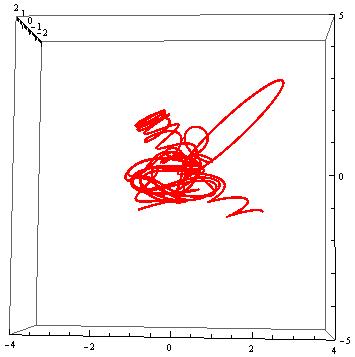} \hfill 
       \includegraphics[width=0.45\textwidth]{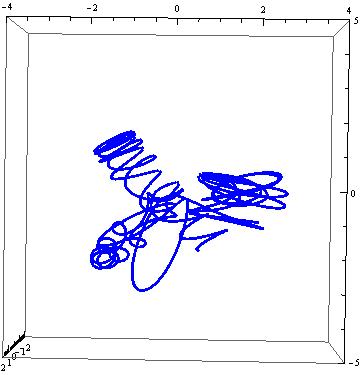}\\}
		\includegraphics[width=0.45\textwidth]{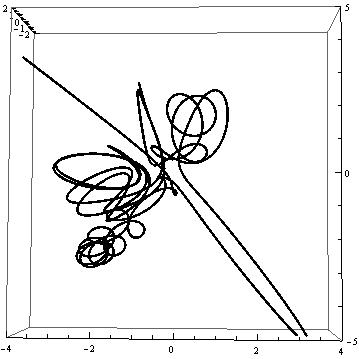}
       } 
 
    \caption{The graphs show the trajectories of the Pythagorean system of three bodies: $m_1$ - in red, $m_2$ - in blue and $m_3$ - in black up to 80 time units obtained with a time step of $10^{-8}$.}
    \label{fig6}
\end{figure*}
\section{Discussion}
The solutions of the problem are deterministic, but depending on the initial conditions adopted, they can still be chaotic, but thanks to the fulfillment of more conditions by the system, as long as the minimum distance between each pair of bodies is defined, the mutual position of bodies can be determined with the accuracy related to the numerical representation of numbers. 
Partial momenta have the additional advantage that their determination uses the force coming from one body, thanks to which expressions having the same order of magnitude are added. This advantage is not available for total momenta, which adds terms from different bodies at each step, which can lead to a rapid loss of precision in the solutions.
Another problem when using total momenta is determining the initial conditions for the momenta relative to the center of mass with high accuracy.
Without sufficient knowledge of the system under consideration, when solving only a differential problem located on the main diagonal of the matrix equation:
$$\frac{d \mathbf{P}}{dt} = \mathbf{F},$$
deterministic determination of the positions of the bodies may turn out to be impossible.
When a body $i$ orbits another body $j$, it is easier to determine this orbital velocity than the velocity of body $i$ relative to the center of mass. Partial momenta are an artificial mathematical object, and as long as the sum of all partial momenta equals the total momentum, they will allow for obtaining the same solutions, provided numerical errors are ignored. However, they can be used to describe the momentum generated by gravitational interactions between a body orbiting another body by identifying the partial momentum with the orbital momentum, which is a much simpler approach than the description with respect to the center of mass.
\section{Conclusions}
The method and schemes presented in this work allow obtaining solutions that are characterized by higher stability than solutions obtained with existing methods used to obtain approximate solutions of Newton's standard equations describing the position and motion of bodies. 
By taking into account the partial momenta generated by each pair of bodies, it is possible to avoid or at least reduce numerical errors occurring when solving equations that take into account only the positions and velocities of the bodies.

The method used can be generalized to the case of relativistic physics or even to all Hamiltonian systems.

\section*{Acknowledgements}
The author thanks Grzegorz Kwiatkowski for his attention to this work, complementary discussions, and consultations.


\end{document}